\magnification\magstep1
\input psfig.tex
\input tole2.sty
\hoffset=5truemm \hsize=5.8truein \osnovni \hfuzz=6pt \vglue 2cm

\centerline{\bf Quantum parallelism in quantum information
processing}

\medskip

\centerline{Miroljub Dugi\' c$^1$ and Milan M. \' Cirkovi\'
c$^{2}$}

\medskip

\centerline{\it $^1$Department of Physics, Faculty of Science}

\centerline{\it P.O.Box 60, 34 000 Kragujevac, Yugoslavia}

\centerline {E-mail: {\tt dugic@knez.uis.kg.ac.yu}}

\medskip

\centerline{\it $^2$Astronomical Observatory, Volgina 7, 11160
Belgrade, Yugoslavia}

\centerline {E-mail: {\tt mcirkovic@aob.aob.bg.ac.yu}}

\bigskip

{\bf Key words:} quantum operations, fidelity, nonideal
measurement, quantum parallelism

\bigskip

{\bf Abstract:} We investigate distinguishability (measured by
the {\it fidelity\/}) of the initial and the final state of a
qubit, which is an object of the so-called nonideal quantum
measurement of the first kind.  We show that the fidelity of a
nonideal measurement can be greater than the fidelity of the
corresponding ideal measurement. This result is somewhat
counterintuitive, and can be traced back to the {\it quantum
parallelism\/} in quantum operations, in analogy with the quantum
parallelism manifested in the quantum computing theory. In
particular, while the quantum parallelism in quantum computing
underlies efficient quantum algorithms, the quantum parallelism
in quantum information theory underlies this, classically
unexpected, increase of the fidelity.

\bigskip

{\bf 1. Introduction: fidelity and quantum measurement}

\bigskip
The sensitivity of quantum systems to various interactions with
its environment resulting in different kinds of the "quantum
operations" on the actual system is one of the major challenging
problems for the realization of quantum computers [1, 2]. Needless
to say, the efforts undertaken in this regard should, hopefully,
make implementation of the error-correction strategies and methods
easier and more efficient [3-6].

The "quantum operations" [7] generally result in an uncontrollable
change of a qubit's state which is characterized by the decrease
in {\it fidelity\/} [7, 8]. The latter is a useful measure of
distinguishability of the initial and the final qubit's states,
albeit not representing a metric in the qubit's Hilbert state
space. More precisely, fidelity is defined by the following
expression [7, 8]:
$$F(\hat \rho, \hat \sigma) = F(\hat \sigma, \hat \rho) =
{\rm Tr} \sqrt{\hat \rho^{1/2} \hat \sigma \hat \rho^{1/2}}.
\eqno(1)$$

\noindent It equals unity if and only if $\hat \sigma = \hat
\rho$, while it equals zero for the orthogonal initial and final
states (since ${\rm Tr} (\hat \rho \hat \sigma) = 0$). In general,
the fidelity satisfies $0 \le F(\hat \sigma, \hat \rho) \le 1$.
Above, $\hat \sigma$ and $\hat \rho$ represent the initial and
the final state of the qubit. From now on, we distinguish between
the "pure" and the "mixed" quantum state [9, 12], referring to
them as to the {\it state vector\/} and {\it state operator\/}
(statistical operator), respectively. There is also an alternative
characterization of fidelity [8] which proves to be equivalent
with Eq.\ (1).

Here we report on the observation that fidelity of the so-called
{\it nonideal quantum measurements of the first kind\/} can
result in the {\it fidelity increase\/} relative to fidelity of
the corresponding ideal measurements. This result is
counterintuitive, for the simple reason that---relative to the
ideal measurements [9]---the nonideal measurements bear {\it
unavoidable uncertainty (ignorance)\/} in the final state operator
[10, 11]. Needless to say, it is {\it our classical intuition\/}
which tacitly assumes that lack of information on the system's
(qubit's) state should imply decrease of fidelity relative to the
situation(s) in which there is no uncertainty. Simultaneously, we
classically expect the entropy increase to be manifested with the
decrease of fidelity [15]. We hereby show that the rather
unexpected increase of fidelity can be traced back to the {\it
quantum parallelism\/} in quantum information processing, in full
analogy with the quantum parallelism as defined in the quantum
computing theory.

In Section 2 we give precise definition of the nonideal
measurement [10--12], as well as a precise formulation of the task
to be performed. In Section 3 we show that the nonideal
measurements can lead to the fidelity increase. Section 4
contains discussion of this and related non-classical phenomena,
while the conclusions are given in Section 5.

\bigskip

{\bf 2. Nonideal quantum measurements}

\bigskip

A "quantum operation" is defined as the map of an arbitrary
("pure" or "mixed") initial state $\hat \sigma$ [7]:
$$E: \hat \sigma \to \sum_n \hat A_n \hat \sigma \hat
A_n^{\dag}, \eqno(2)$$

\noindent
where the $\hat A_n$ are system operators which satisfy the
completeness relation $\sum_n$ $\hat A_n$ $\hat A_n^{\dag} =
\hat I$; conversely, any map of this form is a quantum
operation.

As a special kind of quantum operations appear the so-called
ideal quantum measurements of the first kind [9, 12], for which
Eq. (2) reads:
$$E: \hat \sigma \to \sum_n \hat P_n
\hat \sigma \hat P_n, \eqno (3)$$

\noindent
where the orthogonal
projectors $\hat P_n$ represent the eigenprojectors of the
measured observable, satisfying $\sum_n \hat P_n = \hat I$.

Physically, the right-hand side of Eq.~(3)---which always
describes a "mixed" state represented by a statistical operator
(state operator: some $\hat \sigma '$ ($\hat{\sigma '}^2 \neq \hat
{\sigma '}$))---can be interpreted as the final state of an {\it
ensemble\/} (of the objects of measurement), which was the object
of a "nonselective" quantum measurement [9, 12] or of a
measurement with the result of measurement ignored. In older
terminology, the latter refers to a "selective" measurement with
the measurement result "unread".

In the quantum information (and computation) issues, the quantum
measurement processes prove to be of substantial importance;
e.g., as the (intermediate or the final) steps in quantum
computing algorithms, as the procedures of preparation of the
qubits' states, or as a formal analogue of the process of
decoherence [13, 14], as well as in some quantum information
protocols. However, as was first pointed out by Wigner [10], and
later elaborated by Araki and Yanase [11], {\it realistic\/}
quantum measurements usually suffer from {\it unavoidable
errors}, i.e.\ from unavoidable uncertainty in the final state of
the measured object. Let us put this notion in the mathematical
terms.

First, without a loss of generality, consider the ideal quantum
measurement of the observable $\hat S_{z}$---the $z$-component of
"spin" (qubit). The ideal measurement of $\hat S_{z}$ is
presented by [9]:
$$\hat U \vert \uparrow\rangle \vert\chi \rangle
= \vert \uparrow\rangle \vert\ + \rangle, \eqno(4)$$ \noindent
where $\hat U$ represents the unitary (Schr\"odinger) evolution in
time of  the combined system "object (qubit) + apparatus (Q+A)",
the initial state vector $\vert \uparrow\rangle$ is an eigenstate
of $\hat S_{z}$ for the value $+\hbar/2$, for arbitrary initial
state vector $\vert \chi \rangle$ of the apparatus, and we omit
the unnecessary symbol of the tensor product. Similarly, for the
initial state vector of the object $\vert \downarrow\rangle$ which
is the eigenstate of $\hat S_{z}$ for the value $- \hbar/2$, the
ideal measurement is defined as:
$$\hat U \vert \downarrow\rangle \vert\chi \rangle
= \vert \downarrow\rangle \vert\ - \rangle, \eqno(5)$$
\noindent
while $\langle + \vert - \rangle = 0$.

However, as it was emphasized by Wigner [10], these expressions
refer directly {\it only\/} to the quantum measurements of the
{\it constants of motion}. Following Wigner [10], Araki and Yanase
[11] showed that quantum measurements of observables which are
not the constants of motion are possible, but only with the
limited accuracy. Actually, for the {\it nonideal quantum
measurement\/} of $\hat S_z$ one obtains (we introduce
normalization factors in the original expressions [11]):
$$\hat U \vert \uparrow \rangle \vert
\chi \rangle = (1 - \epsilon^2_{\uparrow})^{1/2}
\vert \uparrow\rangle \vert + \rangle
+ \epsilon_{\uparrow} \vert \downarrow\rangle
\vert -\rangle, \eqno(6)$$
$$\hat U \vert \downarrow \rangle \vert
\chi \rangle = (1 - \epsilon^2_{\downarrow})^{1/2} \vert
\downarrow\rangle \vert - \rangle + \epsilon_{\downarrow} \vert
\uparrow\rangle \vert +\rangle, \eqno(7)$$ \noindent for the
cases considered above, respectively. Subsequently, Yanase [11]
was able to show that:
$$\epsilon_{\uparrow}^2 + \epsilon_{\downarrow}^2
= \epsilon^2 \ge (8 \Vert \hat M_{x}\Vert^2)^{-1} \eqno(8)$$
\noindent
where $\hat M_{x}$ represents an additive constant of
motion of the apparatus; note that, as the apparatus becomes more
macroscopic, the lower bound of $\epsilon$ becomes smaller [11].

Now, relative to the ideal measurements presented by Eqs.\ (4) and
(5), nonideal measurements introduce an {\it unavoidable error\/}
$\epsilon$ in knowing the value of the measured quantity.
Actually, as it directly follows from, e.g., Eq.\ (6), the
ensemble final state operator reads:
$$\hat \rho' = tr_A [\hat U \vert \uparrow\rangle \langle
\uparrow\vert \otimes \vert \chi\rangle \langle \chi\vert \hat
U^{\dag}] = (1 - \epsilon^2/2) \vert \uparrow\rangle \langle
\uparrow\vert + \epsilon^2/2 \vert \downarrow\rangle \langle
\downarrow\vert, \eqno(9)$$ \noindent where we have used the
equality $\epsilon_{\uparrow} = - \epsilon_{\downarrow} =
\epsilon/\sqrt{2}$, which follows from the expression in (8) and
from the normalization condition $\langle \uparrow\vert \langle
\chi\vert \hat U \hat U^{\dag} \vert \downarrow\rangle \vert
\chi\rangle = 0$. With "$tr_A$" we denote the "tracing out" of
the apparatus degrees of freedom. Physically, this error is
substantial: albeit the ensemble of objects is in an eigenstate of
the measured observable (the eigenvalue is $\hbar/2$), the
measurement leads to the opposite (wrong) result, giving the
value $- \hbar/2$ with nonzero probability $\epsilon^2/2$. From
the {\it information-theoretic point of view}, this error
introduces {\it unavoidable uncertainty (ignorance)\/} about the
final state (i.e.\ instead of the state vector $\vert
\uparrow\rangle$, the final state is the state operator given by
Eq.\ (9)), which is {\it classically expected to give rise to a
fidelity decrease}, relative to the fidelity of the ideal
measurement (where $\epsilon = 0$).

Surprisingly enough, we will show that this is not necessarily
the case. Actually, we will show that fidelity of the nonideal
quantum measurement can {\it increase}, thus constituting a
counterintuitive result: having less control imposed on the final
state, one obtains better fidelity of the operation considered.

\bigskip

{\bf 3. Nonideal measurements can increase the fidelity}

\bigskip

Let us first consider the cases studied in Section 2. And let us
introduce the indices "{\it id\/}" and "{\it nonid\/}" for the
ideal and nonideal measurements, respectively.

From the expressions of Eqs.\ (4) and (5) for the ideal
measurement of $\hat S_z$ it obviously follows that the final and
the initial state operators are equal, $\hat \rho = \hat \sigma$,
which gives rise---for both expressions (4) and (5)---to the
maximum fidelity $F(\hat \sigma, \hat \sigma) = 1$. However, for
the nonideal measurements presented by Eqs. (6) and (7), the
initial and the final state operators are not equal anymore.
Actually, e.g., from Eq.\ (6), it follows that the final state
operator is given by Eq.\ (9), thus giving rise to the fidelity
of the measurement:
$$F_{nonid} \equiv F(\hat \rho_{nonid}, \hat \sigma) =
\langle \uparrow \vert \hat \rho_{nonid} \vert \uparrow
\rangle^{1/2} = (1 - \epsilon^2)^{1/2}, \eqno (10)$$ \noindent
where we have used the symmetry property of the fidelity (cf.\
Eq.\ (1)) and the fact that the initial state is the "pure"
quantum state, $\hat \sigma \equiv \vert \uparrow\rangle \langle
\uparrow\vert$. Now, for the difference of the two fidelities we
obtain:
$$\Delta F \equiv F_{id} - F_{nonid} \cong
\epsilon^2/2 > 0, \eqno (11)$$

\noindent
as one would {\it classically expect}.

Let us now consider arbitrary initial state vector of the qubit:
$$\vert \Psi \rangle = \alpha \vert \uparrow \rangle
+ \beta \vert \downarrow \rangle, \quad
\vert \alpha\vert^2 + \vert \beta\vert^2 = 1, \eqno (12)$$

\noindent
and let us calculate the fidelities of the ideal and
nonideal measurements of $\hat S_z$.

For the ideal measurement, the expressions Eqs.\ (4), (5) give:
$$\hat U \vert \Psi\rangle \vert \chi\rangle =
\alpha \vert \uparrow\rangle \vert + \rangle +
\beta \vert \downarrow\rangle \vert -\rangle,
\eqno(13)$$

\noindent which gives for the state operator of the qubit:
$$\hat \rho'_{id} =
\vert \alpha\vert^2 \vert \uparrow\rangle \langle \uparrow \vert
+ \vert \beta\vert^2 \vert \downarrow\rangle \langle
\downarrow\vert. \eqno(14)$$
\noindent
Now, the fidelity of the
ideal measurement reads:
$$F'_{id} = \langle \Psi\vert \hat \rho'_{id}
\vert \Psi\rangle^{1/2} = [1 - 2 \vert \alpha\vert^2
+ 2 (\vert \alpha\vert^2)^2]^{1/2}. \eqno (15)$$

On the other side, for the case of nonideal measurement, after
some simple algebra, in analogy with Eq.\ (13), one obtains:
$$\hat U \vert \Psi\rangle \vert \chi \rangle =
[\alpha(1 - \epsilon_{\uparrow}^2)^{1/2} + \beta
\epsilon_{\downarrow}] \vert \uparrow\rangle \vert + \rangle +
[\beta (1 - \epsilon_{\downarrow}^2)^{1/2} + \alpha
\epsilon_{\uparrow}] \vert \downarrow\rangle \vert - \rangle,
\eqno(16)$$ \noindent which after the tracing out gives for the
state operator of the qubit:
$$\hat \rho'_{nonid} = \vert \alpha (1 - \epsilon^2_{\uparrow})^{1/2}
+ \beta \epsilon_{\downarrow}\vert^2 \vert \uparrow\rangle
\langle \uparrow\vert + \vert \beta (1 -
\epsilon_{\downarrow}^2)^{1/2} + \alpha \epsilon_{\uparrow}
\vert^2 \vert \downarrow\rangle \langle \downarrow\vert.
\eqno(17)$$ \noindent The corresponding fidelity computation
gives:
$$F'_{nonid} = \langle \Psi\vert \hat \rho'_{nonid}
\vert \Psi\rangle^{1/2} =
\{ \vert \alpha\vert^2 \cdot \vert \alpha(1 - \epsilon^2/2)^{1/2} -
2^{-1/2} \beta \epsilon \vert^2 +$$
$$ + \vert \beta \vert^2 \cdot \vert \beta(1 - \epsilon^2/2)^{1/2} +
2^{-1/2} \alpha \epsilon \vert^2 \}^{1/2}. \eqno(18)$$

To simplify the expression Eq.\ (18) we treat the complex numbers
$\alpha$ and $\beta$ as the vectors in plane, so defining the
angle $\theta$ as $\cos \theta = \vec \alpha \cdot \vec \beta /
\vert \alpha\vert \cdot \vert \beta\vert$. Then Eq.\ (18) reads:
$$F'_{nonid} = \{1 - \epsilon^2/2 + 2 (\vert \alpha\vert^2)^2
- 2 (\vert \alpha\vert^2)^2 \epsilon^2 +
2 \vert \alpha\vert^2 \epsilon^2 - 2 \vert \alpha\vert^2 -$$
$$- \epsilon \vert \alpha \vert^3
[8 (1 - \vert \alpha\vert^2) (1 - \epsilon^2/2)]^{1/2} \cos \theta
+
\epsilon \vert \alpha\vert [2(1 - \vert \alpha \vert^2)
(1 - \epsilon^2/2)]^{1/2} \cos\theta\}^{1/2}. \eqno(19)$$

From Eqs.\ (15) and (19) we obtain:
$$F^{'2}_{id} - F^{'2}_{nonid} = \epsilon^2/2
+ 2 (\vert \alpha \vert^2)^2 \epsilon^2
- 2 \vert \alpha\vert^2 \epsilon^2 +$$
$$\epsilon \vert \alpha \vert^3
[8 (1 - \vert \alpha\vert^2) (1 - \epsilon^2/2)]^{1/2} \cos \theta
- \epsilon \vert \alpha\vert [2(1 - \vert \alpha \vert^2) (1 -
\epsilon^2/2)]^{1/2} \cos\theta. \eqno(20)$$
\noindent
Keeping in
mind positivity of the fidelity, the fidelity difference,
$F'_{id} - F'_{nonid}$, is of the same sign as the difference
given by Eq.\ (20).

In Fig.\ 1 we give the plot of $F'^2_{id} - F'^2_{nonid}$ {\it
against} ($\vert\alpha\vert^2, \theta$), for the respecting
intervals $[0,1]$ and $[0, \pi]$, for the two values of
$\epsilon$. As it is obvious from Fig.~1, the shape of the plot is
independent on the values of $\epsilon$, while the maximum
(minimum) value(s) of the difference is of the order of $0.1
\epsilon$ ($10 \epsilon$).

\vfill \eject

{\vskip -2cm}

\psfig{file=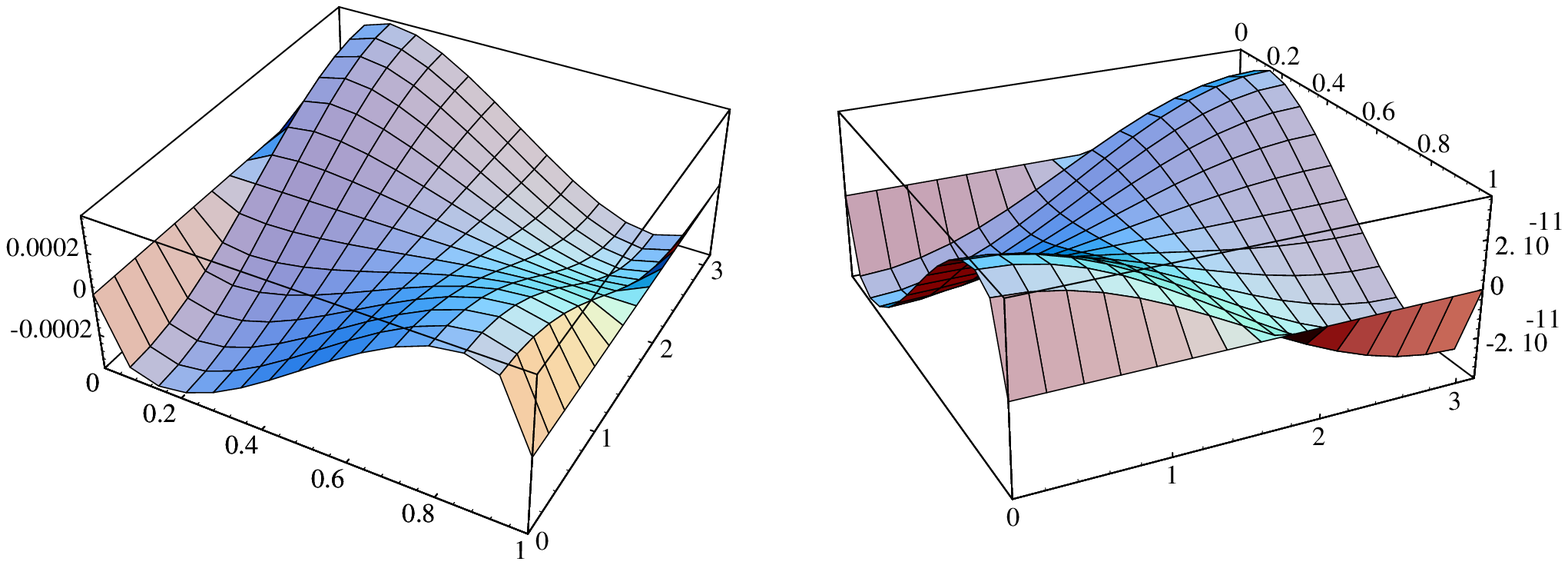,width=12cm,height=14cm}

{\vskip -3cm}
{{\bf Figure 1.} The plot of $F'^2_{id} -
F'^2_{nonid}$ {\it against} $(\vert \alpha\vert^2, \theta)$, for
the respecting intervals $[0, 1]$ and $[0, \pi]$, for the two
values of $\epsilon$ (note the different orientations of the
figures): (a) $\epsilon = 10^{-3}$, and (b) $\epsilon =
10^{-10}$. The values of the variables are in the horizontal
plane. The shape of the plot is independent on the values of
$\epsilon$: In the both cases appears that the given difference
(and also the fidelity difference $F'_{id} - F'_{nonid}$) is {\it
negative} for the following combinations of the intervals:
$\{\vert\alpha\vert^2 \in (0, 0.5) \quad and \quad \theta \in (0,
\pi/2)\}$, as well as for $\{\vert\alpha\vert^2 \in (0.5, 1)
\quad and \quad \theta \in (\pi/2, \pi)\}$. The maximum (minimum)
value of the difference is of the order of $10\epsilon$
($0.1\epsilon$).}

\vfill \eject

From Fig.~1 we conclude that our classical intuition, which is
confirmed by Eq.\ (11), is also confirmed by Fig.\ 1, {\it but
only for special choice\/} of $\vert \alpha \vert^2$ and $\theta$.
Actually, as it is obvious from Fig.~1, one has, for instance:
$$F'_{nonid} > F'_{id}, \vert \alpha\vert^2 \in
(0, 0.5) \quad and \quad \theta \in (0, \pi/2), \eqno(21)$$

\noindent
which challenges our classical intuition.

\bigskip

{\bf 4. Interpretation of the fidelity increase}

\bigskip

The quantum operations (nonideal measurements) presented by Eqs.\
(6) and (7), are "coherently superimposed" in the operation
presented by Eq.\ (16). That is, due to the linearity of the
Schr\"odinger law, the two mutually independent operations in
Eqs.\ (6) and (7) are coherently superimposed in Eq.\ (16). This
gives rise to the fidelity increase as presented by Eq.\ (21).
This observation calls for {\it analogy\/} with the {\it quantum
parallelism\/} as defined in the quantum computation theory [1,
2]. But this interpretation leads to the following question:
whether the fidelity increase can or cannot ever be observed for
the arbitrary initial "mixed state" ("incoherent mixture" of the
initial state vectors $\vert \uparrow \rangle$ and $\vert
\downarrow \rangle$ of the qubit [9, 12]) represented by a state
operator, rather than by a state vector?

To answer this question we consider arbitrary initial "mixed
state" (state operator):
$$\hat \sigma = W_1 \vert \uparrow \rangle \langle \uparrow \vert
+ W_2 \vert \downarrow \rangle \langle\downarrow \vert, \quad W_1
+ W_2 = 1, \eqno (22)$$ \noindent and calculate the fidelities of
both ideal and nonideal measurement of $\hat S_z$; note that for
$W_1 =1$ ($W_2 = 1$) one obtains the case(s) studied above, i.e.\
the expression(s) Eq.\ (4) (Eq.\ (5)).

Again, as it can be easily shown, the ideal measurement of $\hat
S_z$ does not change the initial state of the qubit, thus giving
rise to the maximum fidelity. However, for the nonideal
measurement, after some algebra one obtains for the state
operator of the qubit:
$$\hat \rho''_{nonid} = tr_A [\hat U \hat \sigma
\vert \chi \rangle \langle \chi \vert \hat U^{\dag}]
= [W_1 (1 - \epsilon^2/2) + W_2 \epsilon^2/2]
\vert \uparrow \rangle \langle \uparrow \vert +$$
$$+ [W_2 (1 - \epsilon^2/2) + W_1 \epsilon^2/2]
\vert \downarrow \rangle \langle \downarrow \vert. \eqno (23)$$
\noindent
Therefore, the fidelity in this case reads:
$$F''_{nonid} = tr \{\hat \rho_{nonid}^{' 1/2}
\hat \sigma \hat \rho_{nonid}^{' 1/2}\}^{1/2} =
tr \{\hat \sigma \hat \rho'_{nonid}\}^{1/2} =$$
$$= W_1 \{1 - \epsilon^2/2 + (1 - W_1) \epsilon^2/2W_1\}^{1/2}
+ (1 - W_1) \{1 - \epsilon^2/2 + W_1 \epsilon^2/2(1 - W_1)\}^{1/2}
\le 1, \eqno (24)$$ \noindent where we used $[\hat \sigma, \hat
\rho'_{nonid}] = 0$, and we perceive the last inequality as
obvious, while equality refers only to $W_1 = W_2 = 1/2$. It is
worth emphasizing that for $W_1 = 1$ ($W_2 = 1$), the expression
in Eq.\ (24) reduces to Eq.\ (10).

Therefore, for the state operator $\hat \sigma$, one {\it never
obtains the fidelity increase}, which justifies that the fidelity
increase is {\it ultimately\/} due to the coherent superpositions
of the qubit's states, Eq.\ (12) (and also due to linearity of the
Schr\"odinger law---cf.\ Eq.\ (16)). In other words, the
classically unexpected fidelity increase is caused by the quantum
coherence, i.e.\ is due to the parallel quantum operations in the
coherent mixtures of the qubit's state vectors.

\bigskip

{\bf 5. Discussion}

\bigskip
The main result of this paper is the fidelity increase due to
nonideal quantum measurements, as pointed out by Eq.\ (21). The
interpretation given in the previous Section allows for analogy
with the quantum parallelism as distinguished in the quantum
computing theory. To this end one may note that, while the
quantum parallelism is a general feature of the quantum
computation [1, 2], the fidelity increase we point out refers to
the limited case of the "pure" state, cf.\ Eq.\ (21). Then the
following reasoning may seem plausible: since the fidelity
increase does not refer to arbitrary state vector $\vert
\Psi\rangle$, the interpretation given in Section 4 should be
considered incomplete, if not incorrect. But, we believe this
conclusion to be wrong.

Instead, we maintain that our interpretation is of the general
validity (i.e.\ applicable to any "pure" state of the qubit), {\it
while not requiring\/} the general increase of the fidelity for
nonideal measurements. That is, the operations considered do not
require the fidelity increase {\it per se}. Rather, the fidelity
increase should be considered {\it to point to\/} the quantum
parallelism, being its (classically unexpected) consequence.
Again, we can make analogy with the quantum parallelism in
quantum computing: quantum parallelism in quantum computing does
not {\it a priori\/} guarantee that any quantum-computation
algorithm will be substantially more efficient than its classical
counterpart {\it per se}. However, it is well-known that the
major motivation for research in quantum computing is exactly the
fact that specific implementations of a concrete
quantum-computation algorithm can be shown to be more efficient
than any possible classical analogue. This, we believe, justifies
the {\it full analogy\/} between the quantum parallelism in
quantum computing, and in the quantum-measurement-like quantum
operations (e.g., the decoherence) on a qubit.

The following question may at first sight seem reasonable: does
the fidelity increase can be used for achieving the fidelity
arbitrarily close to unity, by choosing a "sufficiently nonideal"
measurement, i.e.\ by choosing sufficiently big $\epsilon$? But
this question hides a misinterpretation of our result. In the
presumed case one would have, at least as to the lower bound of
$\epsilon$ (cf.\ Eq.\ (8)) to be a reasonable fraction of unity.
However, as the r.h.s.\ of Eq.\ (8) implies, the norm of the
(bounded) observable $\hat M_{x}$ would have to be of the order
of unity, which is physically unacceptable. Actually, such
observable would refer to a {\it microscopic\/} system (here:
apparatus), in contradistinction to the requirement that the
apparatus should be sufficiently macroscopic [9, 11-13]. In other
words, the use of the nonideal measurements in approaching the
equality $F(\hat \sigma, \hat \rho) \cong 1$ would contradict
applicability of the formulae used in the above calculations
(which refer only to the "macroscopic apparatus").

Finally, our considerations bear full generality due to: (i) all
the results concerning the measurements of $\hat S_{z}$ can be
straightforwardly applied to the measurements of arbitrary
observable of a qubit, (ii) the quantum parallelism can be
observed only in comparison of the results (i.e.\ of the
fidelities) for the "coherent" (12) and for the "incoherent" (22)
mixtures  of the {\it same state vectors}. The latter is the
reason we do not calculate fidelity for an arbitrary mixed state
of the qubit. Finally, (iii) the considerations given here can be
straightforwardly extended to account for the $N$-qubit systems.
The latter circumstance can undoubtedly be of some practical
interest in the quantum information research.

\bigskip

{\bf 6. Conclusion}

\bigskip
It is to be expected that lack of information (uncertainty) about
a system's state should be characterized by the fidelity decrease,
relative to the situations in which there is not the uncertainty.
In the context of the quantum measurement process, which is a
special kind of the so-called "quantum operations", this classical
expectation stems that the fidelity of a nonideal quantum
measurement should exhibit decrease, relative to the
corresponding ideal measurement. However, and {\it contrary to
this classical intuition}, we show that the nonideal measurements
can lead to the {\it fidelity increase}. That is, the fidelity of
the nonideal measurements can be greater than the fidelity of the
corresponding ideal measurements. This {\it counterintuitive\/}
result can be traced back to the {\it quantum parallelism\/} in
the quantum information processing, in full analogy with the
quantum parallelism as conventionally discussed in the quantum
computing theory. One may note that, as the quantum parallelism
underlies the efficient quantum computing algorithms, the quantum
parallelism underlies the classically unexpected increase of the
fidelity of the nonideal quantum measurements.

\bigskip

{\bf References:}

\bigskip

\item{[1]}
P.\ W.\ Shor, SIAM J. Comp. {\bf 26} (1997) 1484; L.\ K.\ Grover,
Phys. Rev. Lett. {\bf 79} (1997) 325

\item{[2]}
A. Yu. Kitaev, A. Kh. Shen', M.\ N.\ Vyalyi, {\it "Classical and
Quantum Computation"} (MCCME, CheRo, Moscow, 1999 (in Russian));
D. Aharonov, in {\it Annual Reviews of Computational Phy\-si\-cs},
ed. Dietrich Stauffer (World Scientific, Singapore, 1998) vol. VI

\item{[3]}
P.\ W.\ Shor, Phys. Rev. A {\bf 52} (1995) R2943; A. Ekert and C.
Macchiavello, Phys. Rev. Lett. {\bf 77} (1996) 2585; A.\ M.\
Steane, Phys. Rev. Lett. {\bf 77} (1996) 793; D.\ P.\ DiVincenzo
and P.\ W.\ Shor, Phys. Rev. Lett. {\bf 77} (1996) 3260

\item{[4]}
L. Viola and S. Lloyd, Phys. Rev. A {\bf 58}
(1998) 2733

\item{[5]}
P.\ Zanardi and M.\ Rasetti, Phys. Rev. Lett. {\bf 77} (1997) 3305

\item{[6]}
M.\ Dugi\' c, Quantum Computers and Computing {\bf 1} (2000) 102

\item{[7]}
K.\ Kraus, {\it "States, Effects, and Operations"}
(Springer-Verlag, Berlin, 1983)

\item{[8]}
A.\ Uhlman, Rep. Prog. Math. Physics {\bf 9} (1976) 273

\item{[9]}
J.\ von Neumann, {\it "Mathematical Foundations of Quantum
Mechanics"} (Prin\-ce\-ton University Press, Prin\-ce\-ton, 1955)

\item{[10]}
E.\ P.\ Wigner, Z. Physik {\bf 133} (1952) 101

\item{[11]}
H.\ Araki and M.\ M.\ Yanase, Phys. Rev. {\bf 120} (1960) 622;
M.\ M.\ Yanase, Phys. Rev. {\bf 123} (1961) 666

\item{[12]}
B.\ d'Espagnat, {\it "Conceptual Foundations of Quantum
Mechanics"} (Benjamin, Reading, MA, 1971)

\item{[13]}
W.\ H.\ Zurek, Phys. Rev. D {\bf 26} (1982) 1862; Prog. Theor.
Phys. {\bf 89} (1993) 281

\item{[14]}
M.\ Dugi\' c, Physica Scripta {\bf 53} (1996) 9; Physica Scripta
{\bf 56} (1997) 560

\item{[15]}
S.\ Lloyd, Phys. Rev. A {\bf 39} (1989) 5378

\vfill\eject

\end